\documentclass[pre,a4paper,twocolumn,superscriptaddress,showpacs,nofootinbib,floatfix]{revtex4}

\usepackage{amsmath}
\usepackage{amsfonts}
\usepackage{amssymb}
\usepackage{bm}
\usepackage{latexsym}
\usepackage[dvips]{graphicx,color}
\usepackage{float}

\newcommand{\rsfig}[1]{
  \begin{center}
    \includegraphics*[width=8.5cm]{{#1}}
  \end{center}
}
\newcommand{\rrsfig}[2]{
  \begin{center}
    \includegraphics*[width=0.46\textwidth]{{#1}}
    \hfill{}
    \includegraphics*[width=0.46\textwidth]{{#2}}
  \end{center}
}

\newcommand{\comment}[1]%
{\textsf{\textcolor{red}{#1}}}

\begin{document}
\sloppy

\title{Polymer chain generation for  coarse-grained models using radical-like polymerization}

\author{Fabien~Leonforte}
\email[Email: ]{fabien.leonforte@insa-lyon.fr}
\affiliation{
Universit\'e de Lyon, INSA de Lyon, MATEIS, CNRS, UMR  5510, 69621 Villeurbanne, France}
\author{Michel ~Perez}
\email[Email: ]{michel.perez@insa-lyon.fr}
\affiliation{
Universit\'e de Lyon, INSA de Lyon, MATEIS, CNRS, UMR  5510, 69621 Villeurbanne, France}
\author{Olivier~Lame}
\email[Email: ]{olivier.lame@insa-lyon.fr}
\affiliation{
Universit\'e de Lyon, INSA de Lyon, MATEIS, CNRS, UMR  5510, 69621 Villeurbanne, France}
\author{Jean-Louis Barrat}
\email[Email: ]{jean-louis.barrat@univ-lyon1.fr} \affiliation{
 Universit\'e de Lyon; Univ. Lyon I, Laboratoire de
Physique de la Mati\`ere Condens\'ee et Nanostructures; CNRS, UMR
5586, 69622 Villeurbanne, France}

\begin{abstract}
An innovative method is proposed to generate configurations of
coarse grained models for  polymer melts. This method, largely
inspired by chemical ``radical polymerization'', is divided in three
stages: (i) \emph{nucleation} of radicals (reacting molecules
caching monomers); (ii) \emph{growth} of chains within a solvent of
monomers; (iii) \emph{termination}: annihilation of radicals and
removal of residual monomers. The main interest of this method is
that relaxation is performed  as chains are generated. Pure mono and
poly-disperse polymers melts  are generated and compared to the
configurations generated by the \emph{Push Off}  method from Auhl
\emph{et al.}\cite{Auhl}. A detailed study of the static properties
(gyration radius, mean square internal distance, entanglement
length) confirms that the \emph{radical-like} polymerization
technics is suitable to generate equilibrated melts. The method is flexible, and can be adapted  to generate nano-structured polymers,
namely diblock and triblock copolymers.
\end{abstract}


\maketitle

\section{Introduction}
\label{sec:intro}

Molecular simulation is becoming an increasingly popular tool for
the investigation of mechanical and thermo-mechanical properties of
polymer materials. It can be applied to investigate the properties
of homopolymer systems as well as to nanostructured copolymers or
polymer based nanocomposites, and to gain a microscopic
understanding of the  properties of these technologically important
materials.

The main issue is to understand relations between polymer
nanostructure and, in particular, mechanical  properties. In order
to bridge the gap between micro and macro scales, coarse grained
molecular dynamics, where each "bead" represents several monomers,
are becoming a standard tool. They allow for an investigation of
qualitative and quantitative issues not directly accessible to
experiments, while remaining affordable in terms of computational
costs.

Investigating structure-property relations in  polymeric systems,
however,  requires the preparation of equilibrated melts with long
and entangled chains. Above the glass transition, equilibrium can,
in principle, be achieved using long Molecular Dynamics (MD) or
Monte Carlo (MC) simulations. The situation gets difficult for long
chains with relaxation times that can soon exceed the typical
simulation duration of a few nanoseconds, and for nanostructured
polymers (e.g. block copolymers), where equilibration times, even
for short chains, are too long to use MD or MC to produce
equilibrated melts.

For long polymer chains, hybrid methods combining MD and MC  in
particular the so called "double bridging" algorithm
\cite{Theodorou}, have been used to generate well equilibrated
melts. These algorithms, apart from their technical complexity, are
not particularly well suited for extension to more complex
architectures.

The objective of our contribution is to propose an innovative method
for polymer chain generation, (i) based on a realistic approach
close to radical polymerisation\cite{Rigby,Khare}; (ii) particularly
adapted to generate non linear architectures  (branched polymers,
star polymers, copolymers,...) and/or polydisperse chains; (iii)
providing equilibrated melts.

This method, called \emph{``radical-like polymerization''} will be
tested on different system types (mono- and poly-disperse
homopolymers). It will be also compared to more classical \emph{Push
Off} methods \cite{Auhl,Kremer}, which are based on a two steps
process: (i) random gaussian chain generation and (ii)
equilibration. Systems resulting from step (i) are usually quite far
from equilibrium as chains interactions are not taken into account,
requiring thus long equilibration times (step (ii)).

The main idea of \emph{radical-like polymerisation}, is that chains
are partially relaxed \textbf{simultaneously} while polymerization
is achieved.

The manuscript is organized as follows. Section \ref{sec:technical}
describes the method. In section \ref{sec:results}, we apply the
method to several types of homopolymer melts, and show how it can be
tuned to obtained well equilibrated melts at a relatively low
computational cost.
 Finally, we point out that the \emph{radical-like
polymerization method} is suitable for simulating block copolymers,
and give a preliminary illustration of this in section
\ref{sec:copo}.

\section{Description of systems, and methodology}
\label{sec:technical}

Our simulations are carried out  for a well established
coarse-grained model \cite{Kremer} in which the polymer is treated
as a chain of $N = \sum_{\alpha} N_{\alpha}$ beads (where $\alpha$
denotes the species for block copolymers), which we refer to as
monomers, of mass $m=1$ connected by a spring to form a linear
chain. The beads interact with a classical Lennard-Jones
interaction:

\begin{equation}\label{LJpot}
  \mathrm{U^{\alpha\beta}_{LJ}(r)} = \left\{ \begin{array}{ll}
    4\epsilon_{\alpha\beta}\left[\left(\sigma_{\alpha\beta}/r\right)^{12} - \left(\sigma_{\alpha\beta}/r\right)^6\right]
    &  \mbox{, $r\le r_c$}\\
    0 &\mbox{, $r\ge r_c$}
  \end{array}
  \right.
\end{equation}

\noindent where the cutoff distance $r_c = 2.5\sigma_{\alpha\beta}$.
In addition to \eqref{LJpot}, adjacent monomers along the chains are
coupled through the well known anharmonic Finite Extensible
Nonlinear Elastic potential (FENE):

\begin{equation}\label{FENEpot}
  \mathrm{U_{FENE}(r)} = \left\{ \begin{array}{ll}
    -0.5 k R^2_0 \ln{\left(1 - \left(r/R_0\right)^2\right)} &\mbox{, $r\le R_0$}\\
    \infty &\mbox{, $r > R_0$}
  \end{array}
  \right.
\end{equation}

\noindent The parameters are identical to those given in
Ref.~\cite{Kremer}, namely
$k=30\epsilon_{\alpha\beta}/\sigma_{\alpha\beta}^2$ and
$R_0=1.5\sigma_{\alpha\beta}$, chosen so that unphysical bond
crossings and chain breaking are avoided. All quantities will be
expressed in terms of length $\sigma_{11}=\sigma$, energy
$\epsilon_{\alpha\alpha}=\epsilon$ and time
$\tau_{LJ}=\sqrt{m\sigma^2/\epsilon}$.

 Newton's equations of motion
are integrated with velocity-Verlet method and a time step $\delta
t=0.006$. Periodic simulation cells of cubic size $L$ containing $M$
chains of size $N$ where used under a Nos\'e-Hoover barostat,
\emph{i.e.} in the NPT ensemble. The pressure is fixed to $P=
0.5\epsilon/\sigma^3$

\subsection{Radical-like polymerization}\label{sec:technical:A}

\subsubsection{Algorithm}

The \emph{radical-like} polymerization method is directly inspired
by the classical polymerization phenomenon with a protocol based on
three stages:
\begin{itemize}
\item \emph{starting:} a radical (active molecule that interacts with monomers) is created by
an active molecule $A$ ($A\rightarrow P^{\star}$) and interacts with a first monomer $P^{\star} + M\rightarrow PM^{\star}$,

\item \emph{propagation:} the radical captures a new monomer and
moves to the  chain end $PM^{\star} + M\rightarrow PMM^{\star}$

\item \emph{termination:} four main mechanisms of
termination can  usually be  identified in polymerization reactors:
(i) two radicals can annihilate leading to two separated polymer
chains ($PM\ldots M^{\star} + PM\ldots M^{\star}\rightarrow PM\ldots
M + PM\ldots M$) (\emph{disproportination}); (ii) two radicals can
annihilate leading to one  polymer chain ($PM\ldots M^{\star} +
PM\ldots M^{\star}\rightarrow PM\ldots MM\ldots MP$)
(\emph{coupling}); (iii) a radical can be transferred to another
monomer leading to a new growing chain (\emph{transfert}) or
annihilated by some  defect. Radicals can also remain active and
chain growth is stopped only when
 all monomers have been consumed, as in (\emph{living polymerization}).
\end{itemize}

The \emph{radical-like} polymerization process takes place in a
solvent which is represented in our simulations as a Lennard-Jones
liquid of $N_{monom}$ monomers.

Note that the aim of our method is not to model in detail the
polymerization process, but rather to take inspiration from it. As a
reminder, we give in Tab.~\ref{TAB_PARAM} a summary of relevant
parameters fully describing the \emph{radical-like} polymerization
algorithm.

\begin{table}
  \begin{tabular}{ll}
    \hline
    Parameters & Signification \\
    \hline
    $N_{monom}$    & Total number of beads in the simulation box \\
    $M$            & Total number of chains \\
    $N_i$       & Final length for a chain $i$ \\
    $N$            & Desired chain length for isodisperse systems \\
    $p$            & Nucleation probability \\
    $n_{growth}$   & Number of growth steps \\
    $n_{MDSbg}$    & Amount of MD steps between each growth step \\
    $n_{relax}$    & Number of MD steps during relaxation phase \\
    $n_{tot}$      & Total number of MD steps \\
    \hline
  \end{tabular}
  \caption{\label{TAB_PARAM} Relevant parameters used in the \emph{ radical-like polymerization algorithm}.}
\end{table}

\begin{figure}[h!]
  \rsfig{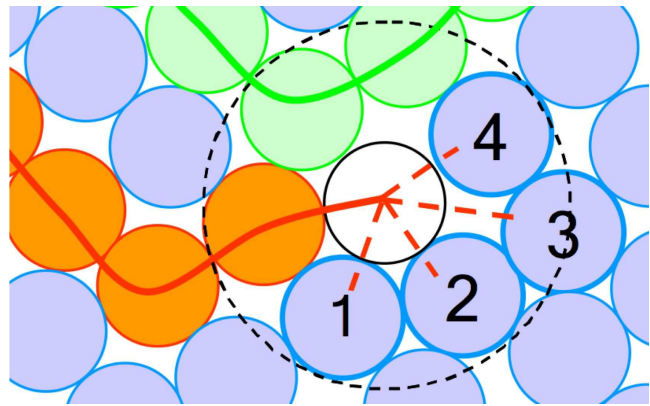}
  \caption{\label{growth_protocol} (Color online). \textbf{Growth} step during
  the \emph{radical-like} polymerization algorithm.
  A radical (white) is randomly assigned  one of its first monomers neighbors
   (blue ones, numbered from $1$ to $4$) to create a new covalent
  bond and increase the local chain length $N_i$.}
\end{figure}

The radical-like algorithm is then divided in four stages:

\begin{enumerate}
  \item \textbf{Nucleation:} each monomer has a probability $p$ to be
  randomly functionalized as a radical. This probability $p$
   controls the  number of chains $M = p\times N_{monom}$ that will eventually be created.

\item \textbf{Growth:} radical (index $i$) randomly chooses one of its
first neighbors still in the monomer state (if any available) to
create a new covalent bond and increase the local chain length $N_i$
of chain $i$. The amount of growth steps $n_{growth}$, defined
initially, controls the maximum chain length
$N_i|_{MAX}=n_{growth}$. Note that this allows us, as mentioned
previously, to mimic the  polydispersity  associated with living
polymerisation. This stage of the process is schematically depicted
on Fig.~\ref{growth_protocol}.

\item \textbf{Relaxation:} this is an essential ingredient   of the method.
Between two successive  growth steps,  a radical is allowed  to
explore
    its neighborhood during $n_{MDSbg}$ MD steps. This is equivalent to let
a chain evolve in the solvent and explore a part of its
conformational phase space \emph{in situ} while polymerization is
taking place, hence permitting  a partial relaxation.

\item \textbf{Termination:} For polydisperse systems the generation procedure
is stopped after a fixed number of growth steps $n_{growth}$. To
produce a monodisperse system, the process is stopped only when each
chain has reached a desired size $N$, whatever the number of growth
steps. Naturally, the time elapsed before termination will depend on
the ratio $(N\times M)/N_{monom}$ which we took near $80\%$.
  Finally the residual monomers (or solvent) are removed
  and the system is relaxed to reach at constant pressure.
\end{enumerate}

In the following we will be only interested in semi-flexible chains
for which no angular potential and bond rigidity are imposed. In
principle, such constraints could be included trough a preferential
choice of the neighbors during the growth step. Three types of
systems were generated using the \emph{radical-like} polymerization
process:

\begin{itemize}
  \item \emph{Cold}: pure \textbf{polydisperse} melt. The polymerization procedure
involve a finite value $n_{growth}$ of growth steps but without
coupling the system to a heat bath by
 imposing $n_{MDSbg}\equiv 0$, thus preventing any relaxation   between  growth steps.
  \item \emph{Hot}: pure \textbf{polydisperse} melt. The number $n_{growth}$ of growth
 steps is also fixed to a finite value,
 but for this kind of polymerization, the system is coupled to a heat bath by fixing a finite
 number $n_{MDSbg}$ of relaxation steps between each growth step, and
 setting MD parameters using a Nos\'e-Hoover barostat with $k_B T=2$ and $P=0.5$.
 For this kind of procedure, the polymerization process is stopped once the number of growth steps
 is reached.
  \item \emph{HotMono}: pure \textbf{isodisperse} melt. For this kind of process, the
  number of growth steps is infinite. Practically, growth stage occur until all
  chains reach the desired size $N$. The system is coupled to a heat bath
   (Nos\'e-Hoover thermostat and barostat with $k_B T=2$ and $P=0.5$) during the relaxation stage.
   $n_{MDSbg}$ MD steps are performed after each growth step.
\end{itemize}

In the next section, these three types of generation processes will
be tested and compared. The monodisperse \emph{HotMono} generation
procedure will also be compared to the more classical
\emph{Push-Off} techniques~\cite{Auhl,Kremer}.

Within the \emph{Push-Off} framework, chains are generated randomly
in the simulation box without considering excluded
volume~\cite{Theodorou}. Thus,  Lennard-Jones interactions for
non-bonded monomers cannot be introduced immediately because chains
spatially overlap. To bypass this difficulty, modified Lennard-Jones
potentials (\emph{Slow Push Off}), or intermediate soft repulsive
potential (\emph{Fast Push Off}) are then introduced, and eventually
replaced by the LJ potential. Due to its relative simplicity, this
method  has been widely used in the literature to generate
monodisperse systems.

We refer to~\cite{Auhl} for details and discussions about FPO
technics. In our implementation, the systems generated with FPO
($M=200$ chains with chain length of $N=200$) are relaxed during
$10^7$ MDS for systems o under Nos\'e-Hoover thermostat ($k_B
T=2.0$) and barostat ($P=0.5$).It has to be noticed that this quite
easy procedure is known to create significant distortions in the
chain statistics on length scales comparable to the tube diameter
\cite{Auhl,Brown,Hoy}, requiring thus relatively long relaxation
times. Consequently chain length is generally limited to $N < 400$.

\subsubsection{parameters}

The values of the parameters used in our  generation processes and
subsequent simulation for  the three types of protocols are
summarized in Tab.~\ref{TAB_SIMU}. For polydisperse systems,
\emph{e.g.} \emph{Cold} and \emph{Hot}, the \emph{min} and
\emph{max} values of the chain length distribution are also quoted
in the same table,   and will be discussed below.

For monodisperse systems, we also studied the influence of the
number of relaxation steps $n_{MDSbg}$ between  growth steps on the
final static properties of the  polymer melt. This parameter  can be
considered as a control parameter for the exploration of
configurational phase space during growth, at a given temperature
and pressure.

\begin{table}[h!]
  \begin{tabular}{|l|c|c|c|c|c|}
    \hline
                  & $\,\,n_{growth}\,\,$ & $\,\,n_{MDSbg}\,\,$ & $\,\,M\,\,$ & $\,\,N\,\,$ & $\langle N\rangle(t\rightarrow\infty)$\\
    \hline\hline
    \emph{Cold}    &   $350$              &   $0$               & $184$       & $[50;344]$  &     $172$\\
    \hline
    \emph{Hot}     &   $6.7\,10^4$        &   $10$              & $215$       & $[56;390]$  &     $226$\\
    \hline
    \emph{HotMono} &   $10^5$             &   $10$              & $215$       & $200$       &     $200$\\
                  &   $10^5$             &   $300$             & $215$       & $200$       &     $200$\\
    \hline
  \end{tabular}
  \caption{\label{TAB_SIMU} Parameters used to simulate the different \emph{ radical-like polymerization}
   processes discussed in text, during
    the generation stage.}
\end{table}

\subsubsection{Structural characterization}\label{sec:technical:B}

Three types structural parameters have been investigated to control
the state of relaxation of polymer melts:
\begin{itemize}
\item the mean gyration radius $\langle r_g\rangle$ defined by:
\begin{equation}
\langle r_g\rangle^2 = \sum_{i=1}^M \frac{ \sum_{j=1}^N\left(\mathrm{r_j^i}-\langle\mathrm{r^i}\rangle \right)^2 /N^i}{M}
\end{equation}
where $\mathrm{r_j^i}$ is the position of the $j$-th atom of the $i$-th chain, $\langle\mathrm{r^i}\rangle$ is the center of mass of chain $i$ and $N^i$ is the size of chain $i$.

\item the Mean Square Internal Distance (MSID) $\langle\mathrm{r}^2\rangle(n)$ is the average
 squared distance between monomers $j$ and $j+n$ of the same chain. It is defined by:

\begin{equation}
\langle\mathrm{r}^2\rangle(n) = \sum_{i=1}^M \frac{ \sum_{j=1}^{N^i-n}\left(\mathrm{r_j^i}- \mathrm{r_{j+n}^i}\right)^2/(N^i-n)}{M}
\label{MSID.eq}
\end{equation}

Note that the MSID $\langle\mathrm{r}^2\rangle(n)$ is a function of $n$ and
$\langle b^2\rangle^{1/2}=\sqrt{\langle\mathrm{r}^2\rangle(1)}$ is the mean bond length.

\item the primitive path: the Primitive Path Analysis is a powerful
tool to investigate the distance between chains entanglements. It is a key parameter,
 that controls the mechanical or rheological properties of the polymer melt.
  Section~\ref{PPA.subsec} will be devoted to the PPA.
\end{itemize}

\section{Results for a homopolymer melt}
\label{sec:results}

\subsection{Dynamics of the polymerization process}

\begin{figure}
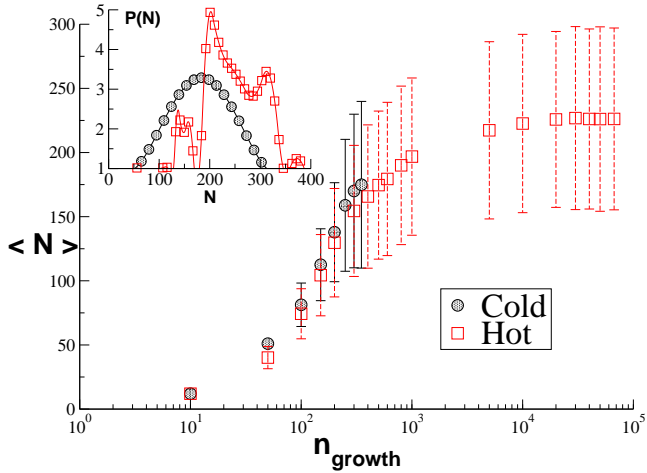

\rsfig{fig1.eps} \caption{\label{dist_Ip_vs_t} (Color online) Mean
chain length size $\langle N\rangle$ (symbols) and polydispersity
index $I_p$ (lines) evolution during polymerization stage versus the
number of growth step, and for the two \emph{Hot} and \emph{Cold}
simulated isodisperse systems. Also plotted is the standard
deviation $\sigma_N$ represented by vertical bars centered on
symbols. \textbf{Inset:} Size distribution $P(N)$ for the same
systems at the end of the generation procedure.}
\end{figure}

A preliminary study is devoted to the growth dynamics of
polydisperse system, namely \emph{Hot} and \emph{Cold} methods. In
Fig.~\ref{dist_Ip_vs_t}, the mean chain length $\langle N \rangle$
is plotted as a function of the number of growth steps preformed
during polymerization. It is worth noting that polydispersity has
spontaneously appeared as a result of the growth process. We observe
that both methods display the same evolution: a rapid increase
followed by a saturation due to the lack of available monomers.
However, the \emph{Cold} procedure is stopped before the \emph{Hot}
one because thermal mobility allows a more efficient exploration of
configurational space by the active  radicals.

The standard deviation $\sigma_N = \sqrt{\langle N^2\rangle -
\langle N\rangle^2}$ is also indicated for both systems with
vertical bars centered on respective symbols.

The final size distributions at the end of the generation procedure
$P(N)$ are plotted for both systems on inset of
Fig.~\ref{dist_Ip_vs_t}. As expected, the peak is shifted towards
larger sizes and is slightly narrower for the \emph{Hot} system.

In our simulations the polydispersity index
$I_p=M_w/M_n$\footnote{$I_p$ is defined be the ratio of $M_w$ the
weight averaged molecular weight and $M_n$ the number averaged
molecular weight}, is  accessible through the ratio $I_p= \langle
N^2\rangle/\langle N\rangle^2$. The final polydispersity index is a
little lower for the \emph{Hot} system (around 1.057) than for the
\emph{cold } system (around 1.103). Again, this is probably due to
thermal mobility which allows smaller chains to find new monomers to
continue the growth.

Our generation procedure, which is very close to living
polymerisation (see section~\ref{sec:technical:A}) leads to
polydispersity indexes that are reasonably close  to experimental
ones resulting from living polymerization (typically of order 1.3),
which gives us confidence in the physical background our the radical
\emph{radical-like} polymerization algorithm. Moreover, it would be
very easy to slightly modify our method to simulate other kind of
polymerization processes which would lead to higher polydispersity.
Experimental values of polydispersity index can reach a value of 10
or more for classical polymers where coupling, transfert or
disproportination are involved (see section~\ref{sec:technical:A}).

\begin{figure}
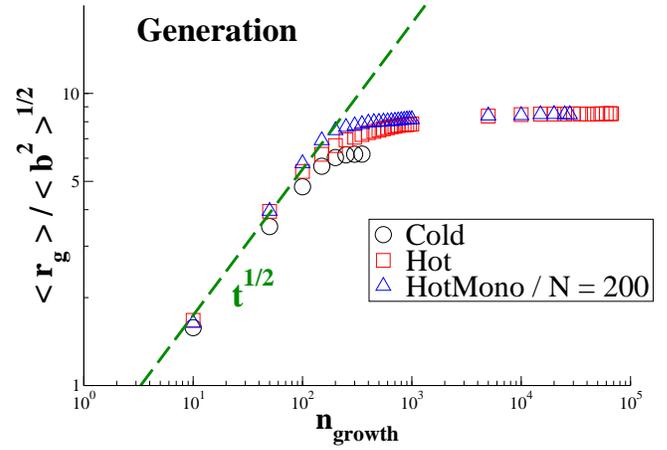

\rsfig{fig2_a.eps} \caption{(Color online) Generation stage:
evolution of the mean gyration radius $\langle r_g(t)\rangle$
normalized by the average bond length $\langle b^2(t)\rangle^{1/2}$
and averaged over all chains. Generation exhibits two distinct
stages: (i) a pure growth stage characterized by a $t^\frac{1}{2}$
growth kinetics; (ii) a saturation stage where gyration radii reach
a plateau value. A value of $n_{MDSbg}=10$ has been used for
\emph{Hot} and \emph{HotMono} methods (see Tab.~\ref{TAB_SIMU}).}
\label{rg_vs_n_growth.fig}
\end{figure}

\begin{figure}
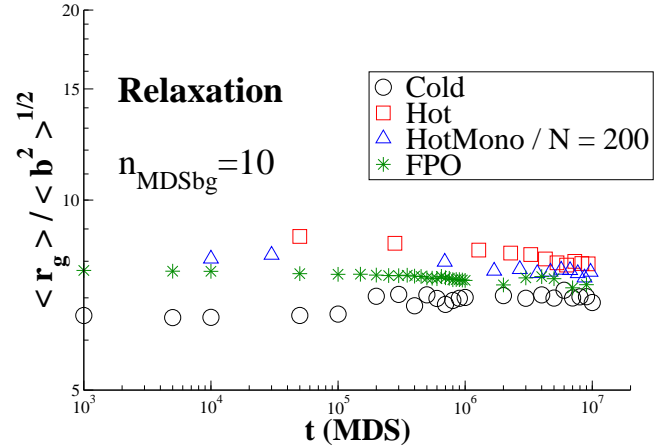

\rsfig{fig2_b.eps} \caption{(Color online) Relaxation stage
(\emph{e.g.} after polymerization): evolution of the mean gyration
radius as a function of the number of MD steps necessary to reach a
total number $n_{tot} = n_{growth} \times n_{MDSbg} + n_{relax} =
10^7$ MD steps. Fast Push Off (FPO) and \emph{HotMono} methods
converge to the same value.} \label{rg_vs_time.fig}
\end{figure}

In order to quantify the evolution of the structural properties of
chains during production runs for the \emph{Hot}, \emph{Cold} and
\emph{HotMono} methods, we also investigated the evolution of the
mean gyration radius $\langle r_g(t)\rangle$ normalized by the mean
bond distance $\langle b^2(t)\rangle^{1/2}$ during the growth
(Fig.~\ref{rg_vs_n_growth.fig}) and relaxation
(Fig.~\ref{rg_vs_time.fig}) stages. Such evolutions are investigated
for the three systems ($n_{MDSbg}=10$ for \emph{Hot} and
\emph{HotMono} during the generation stage - see
Tab.~\ref{TAB_SIMU}).

In figure ~\ref{rg_vs_n_growth.fig}, we observe that the generation
proceeds in  two distinct stages: (i) a pure growth stage
characterized by a $t^\frac{1}{2}$ growth kinetics; (ii) a
saturation stage where gyration radii reach a plateau value. The
power law simply means that during stage (i), each growth step is
successful and lead to an increase of the chain length $N$:
$N\propto n_{growth}$. As $r_g\propto N^\frac{1}{2}$, we obviously
get $r_g\propto n_{growth} ^\frac{1}{2}$.

In figure ~\ref{rg_vs_time.fig}, the time evolution of the mean
radius of gyration for the \emph{Cold}, \emph{Hot}, \emph{HotMono}
and also \emph{Fast Push Off} (FPO) are compared during the
relaxation stage. The  radius of gyration is plotted \emph{versus}
the number of MD steps necessary to reach a total number $n_{tot} =
n_{growth} \times n_{MDSbg} + n_{relax} = 10^7$ MD steps. Final
values of gyration radii depend on mean chain length $N$: the
\emph{Cold} method, which gives the smallest final mean chain length
($N=172$) leads to the smallest mean gyration radius. Then come the
\emph{HotMono} and the FPO methods, which converge logically to the
same gyration radius. Finally, the \emph{Hot} method, which gives
the largest final mean chain length ($N=226$) leads to the largest
mean gyration radius.

\begin{figure}
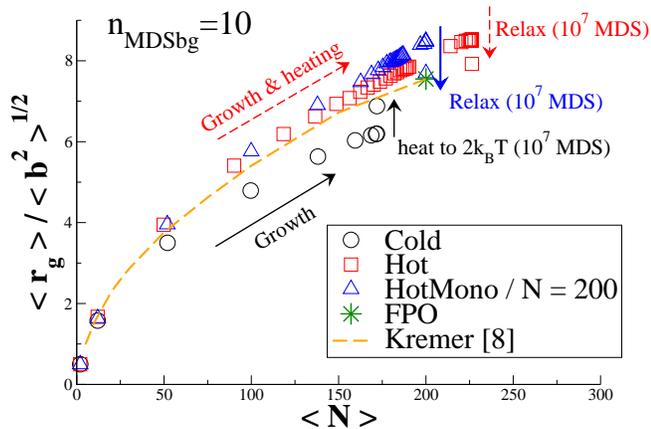

\rsfig{fig3.eps}
\caption{(Color online) Growth and relaxation stages: evolution of the mean
 gyration radius as a function of the mean chain size during growth (curves)
 and relaxation (vertical arrows) stages. Data from Kremer\cite{Kremer} are
 also represented. They predict a $N(t)^{1/2}$ dependance. After the removal
 of remaining monomers and $10^7$ MD relaxation steps, all generation technics
  are in very good agreement with Kremer's results.}
\label{N_vs_t.fig}
\end{figure}

In order to investigate the  evolution of the chain size as a
function of chain length during the growth and relaxation stages for
all polymerization methods, we plotted $\langle
r_g(t)\rangle/\langle b^2(t)\rangle^{1/2}$ \emph{versus} $\langle N
\rangle$ on Fig.~\ref{N_vs_t.fig}. In this figure, relaxation
process (at constant $N$) is represented by vertical arrows. We also
plotted in this figure data from Kremer\cite{Kremer} resulting from
long time equilibration, which predict a $N^{1/2}$ dependance.

After the removal of remaining monomers and $10^7$ MD relaxation
steps, all generation methods (\emph{Cold}, \emph{Hot} and
\emph{HotMono}) are in very good agreement with Kremer's target
function $\langle r_g(t)\rangle/\langle b^2(t)\rangle^{1/2}$
\emph{versus} $\langle N \rangle$.

However, with \emph{Cold}, \emph{Hot} or \emph{HotMono}) (with
$n_{MDSbG}=10$), it seems that relatively long relaxation times (up
to $10^7$ MD steps) are necessary to reach Kremer's target function.
Therefore, in what follows, the effect of the number of MD steps
between each growth step ($n_{MDSbG}$) will be investigated.

\begin{figure}
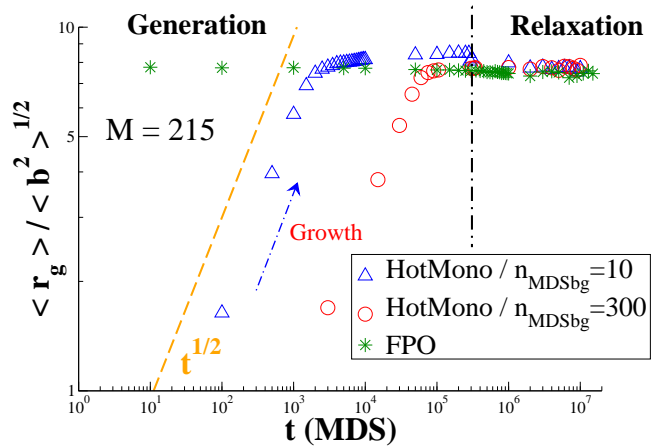

\rsfig{fig4.eps} \caption{(Color online) Evolution of the mean
radius of gyration as a function of time (in MD steps) during growth
and relaxation stages: generation of $M=215$ chains
 of expected length $N=200$ at $k_B T=2$ and $P=0.5$. Two different values of
  $n_{MDSbG}$ (the number of MD steps between each growth step) are compared.
  A larger value of $n_{MDSbG}$ slows down the growth kinetics, but leads to
  better equilibrated systems once growth is completed. For $n_{MDSbG}=300$,
  no equilibration stage is required to reach the mean
   radius of gyration obtained with the Fast Push Off (FPO) method.}
\label{rg_vs_t_MDSbg.fig}
\end{figure}

In Fig.~\ref{rg_vs_t_MDSbg.fig}, the mean normalized  radius of
gyration is plotted \emph{versus} simulation time for the generation
of $M=215$ chains of expected length $N=200$ at $k_B T=2$ and
$P=0.5$. Two different values of $n_{MDSbg}$ are investigated:
$n_{MDSbg}=10$ and $n_{MDSbg} = 300$. It can be observed that a
larger value of $n_{MDSbG}$ slows down the growth kinetics, but
leads to better equilibrated systems once growth is completed. For
$n_{MDSbG}=300$, no equilibration MD steps are required to reach the
 radius of gyration  obtained with the Fast Push Off (FPO) method.

This shows that the  chains generated here reach their equilibrium
structure more rapidly for the protocole that spends more time
during the growth stage\footnote{``To win a race, the swiftness of a
dart, avails not without a timely start...'' \emph{The Hare and the
Tortoise}, Book VI, Jean de La Fontaine.}, thus pointing out the
main interest of this algorithm: \emph{i.e.} equilibration is
occurring during generation, provided an appropriate compromise for
the number of MD steps between growth steps is chosen.

\subsection{Comparison of chains structure for \emph{HotMono} and FPO methods}

The structure of a polymer melt can be characterized by a wide
variety of static or dynamic interchain and intrachain correlation
functions \cite{Kremer,Auhl,Wittmer,Flory,deGennes,Vladkov} which
are more or less sensitive to the artifacts introduced by the
preparation procedure and which equilibrate on different time
scales. One may note that for fully flexible chains simulated in our
model (only FENE + LJ interactions), \emph{i.e.} without torsional
barrier and bending stiffness potentials, the local monomer packing
relaxes quickly, while deviations of chain conformations on large
scale require large times to equilibrate.

To validate our generation methods according to more ``classical''
techniques, we will be interested, in the following, by a measure of
internal chain conformation, namely the Mean-Square Internal
Distance (MSID) $\langle\mathrm{r}^2\rangle(n)$. This function,
defined in Eq.~\eqref{MSID.eq} above, gives  the average squared
distance between two  monomers belonging to the same chain, and
separated by a subchain of $n$ monomers.

\begin{figure}
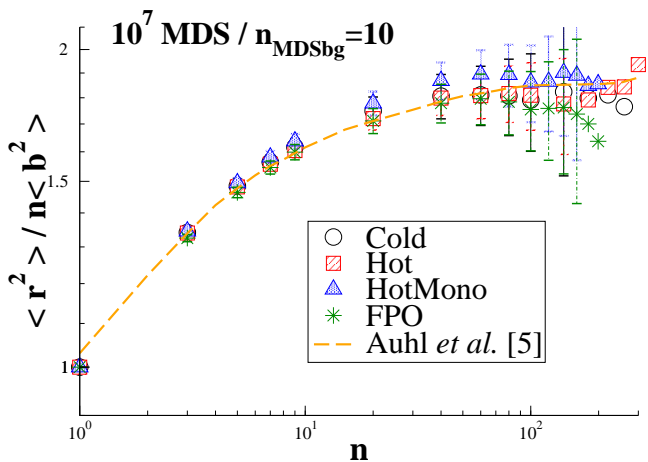

\rsfig{fig5.eps}
\caption{(Color online) Mean square internal distance (MSID) of
generated melts measured after from long MD runs ($10^7$ MD steps).
 All the simulated systems (\emph{Cold}, \emph{Hot}, \emph{HotMono} and FPO
 are compared to the target function of Auhl\cite{Auhl}. Error bars are calculated
 using standard error function on statistical samples. All  methods lead to well equilibrated melts.}
\label{msid_vs_Method_10MDSbg}
\end{figure}

The MSID is shown in Fig.~\ref{msid_vs_Method_10MDSbg} for all
simulated systems after the total number of MD steps $n_{tot} =
n_{growth} \times n_{MDSbg} + n_{relax} = 10^7$ MD steps.
\emph{Cold}, \emph{Hot}, \emph{HotMono} (with  $n_{MDSbg}=10$), and
also \emph{Fast Push Off} (FPO) seem to converge to the same
configuration since they all fit nicely with the ``target function''
defined by Auhl\cite{Auhl} as the signature of well equilibrated
melts.

Error bars  in Fig.~\ref{msid_vs_Method_10MDSbg} are estimated using
the standard error function that includes the number of subset
events taken into account to compute the MSID. As $n$ reaches chain
length $N$ ($n\rightarrow N$),  less and less pairs of monomers are
included in the statistics, leading to large error bars for large
$n$. Hence, error bars for large $n$ have not been represented. We
thus consider that the values obtained for large $n$ are not
statistically significant.

Once again, all our generation methods (\emph{Cold}, \emph{Hot} and
\emph{HotMono}) lead to well equilibrated melts (according to the
MSID criterion) after (i) generation, (ii) removal of remaining
monomers and (iii) $10^7$ MD relaxation steps.

\begin{figure}
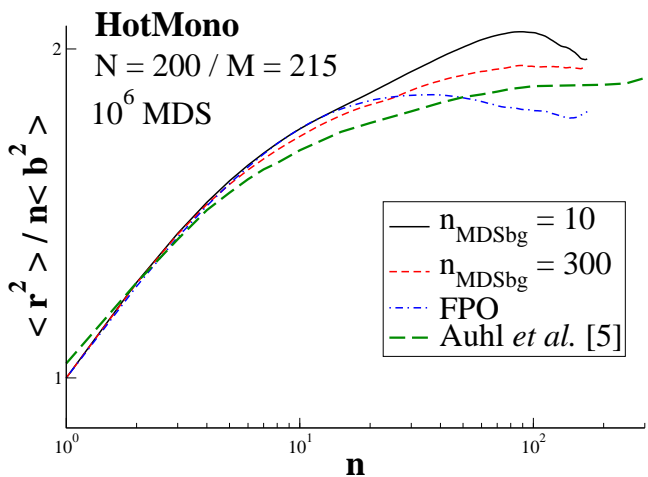

\rsfig{fig6_a.eps} \caption{(Color online) Mean square internal
distance (MSID) of  mono-disperse melts. The effect of the number of
MD steps between each growth step is studied. A larger value of
$n_{MDSbG}$ leads to better equilibrated systems, whom which MSID
fits nicely with FPO and the target function of Auhl\cite{Auhl}.}
\label{MSID_vs_MDSbg.fig}
\end{figure}

As far as the radius of gyration  is concerned, we showed that the
number of MD steps between successive  growth steps ($n_{MDSbG}$)
has an effect on the final structure of the melt. Indeed, the number
($n_{MDSbg}$) of relaxation steps between  growth steps can be view
as a relaxation process for chains during the polymerization stage.
Therefore, the MSID of mono-disperse melts has been investigated for
$n_{MDSbg}=10$ and $n_{MDSbg}=300$.

On Fig.~\ref{MSID_vs_MDSbg.fig}, MSID resulting from \emph{HotMono}
generation (with $n_{MDSbg}=10$  and $n_{MDSbg}=300$) are compared
with MSID resulting from FPO generation and the target function of
Auhl\cite{Auhl}. For all systems, an equilibration stage of $n_{tot}
= n_{MDSbg}\times n_{growth} + n_{relax} = 10^6$ MD steps after
generation has been performed. Despite this relatively low
equilibration time, it can be observed in
Fig.~\ref{MSID_vs_MDSbg.fig} that the \emph{HotMono} generation
method with 300 MD steps between each growth step leads to
relatively well equilibrated systems, even slightly better than FPO
method. This corroborates previous results from
Fig.~\ref{rg_vs_t_MDSbg.fig}, and points out, once again, the main
interest of this \emph{radical-like} generation method: relaxation
takes place while generation is performed.

\subsection{Primitive Path Analysis}
\label{PPA.subsec}

Entanglements between chains are  an important topological feature,
that controls many dynamical properties of polymer melts. A
practical tool for characterizing entanglements is the Primitive
Path Analysis (PPA), which will be the object of this section.

Proposed by Everaers\cite{Everaers} with the  aim of constructing a
real space representation of   de Gennes' tube model, the PPA
technique is an interesting tool for obtaining informations about
the density of entanglements which has not been accessible through
other theoretical or direct experimental measurements.

Recently Hoy and Robbins \cite{Hoy} applied this technique to
quantify the effect of the generation procedure and of the
relaxation for  two types of generation methods, namely the FPO
system and the Double-Bridging\cite{Auhl} relaxation technique.
Following their idea, we apply this to our different
\emph{radical-like} generation methods, first focusing on the
comparison between \emph{HotMono} and FPO method.

The principle of PPA is the following:

\begin{itemize}
\item[(i)] We start with any given configuration, during the growth or in the final state,
after or before  the relaxation step.

\item[(ii)] The two chains ends are kept fixed, while the intra-chain pair
 interaction (covalent bonds) are shifted to get their minimum energy at
 a zero distance while increasing the bond tension in Eq.~\eqref{FENEpot}
to $k=100$;

\item[(iii)] To prevent chain crossing\cite{Sukumaran}, weak bonds lengths have been monitored and limited
 to $1.2\sigma$.

\item[(iv)] The system is then equilibrated using a Conjugate Gradient
algorithm in order to minimize its potential energy and reach a
local minimum.

\item[(v)] The contour length of the primitive
path $L_{pp}$ is then the total length of the chain
(the sum of all straight primitive path segments length).
\end{itemize}

If no entanglement exists between chains, $L_{pp}$ should be equal
to their end-to-end distance $r_{ee}$. The presence of entanglements
leads to $L_{pp} > r_{ee}$ with a typical Kuhn length $a_{pp} =
\langle r^2_{ee} \rangle / L_{pp}$ and an average bond length
$\langle b_{pp} \rangle = L_{pp}/N$. The number of monomers in
straight primitive path segments is then given by:

\begin{equation}\label{Npp_N}
  N_{pp}(N) = \frac{a_{pp}}{\langle b_{pp}\rangle} = \frac{N\langle R^2_{ee}\rangle}{L^2_{pp}}
\end{equation}

For short chains without any entanglements, the primitive path
length equal end-to-end distance leading to $N_{pp}=N$. When chain
lengths are comparable to the entanglement length,   $N_{pp} < N_e$,
$N_e$ being the real entanglement value. For sufficiently long
chains, i.e. $N
> 2N_e$, several entanglements per chains exist, and
$N_{pp}(N)=N_e$.

\begin{figure}
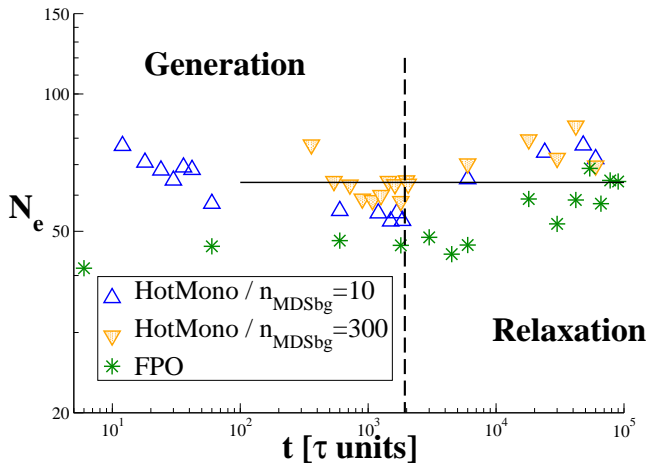

\rsfig{fig7_a.eps} \caption{(Color online) Evolution of the number
$N_e(t)$ of monomers in straight primitive path segments along
simulation times for isodisperse systems \emph{HotMono} and FPO.
Calculation were performed both during the generation (polydisperse)
phase and the relaxation (isodisperse) phase separated by the
vertical dashed line. Units of time are in $\tau$ units, \emph{i.e.}
$n_{tot}\times \delta t$. The horizontal line gives value for $N_e$
from Sukumaran\cite{Sukumaran}.} \label{ppa_homo.fig}
\end{figure}

\begin{figure}
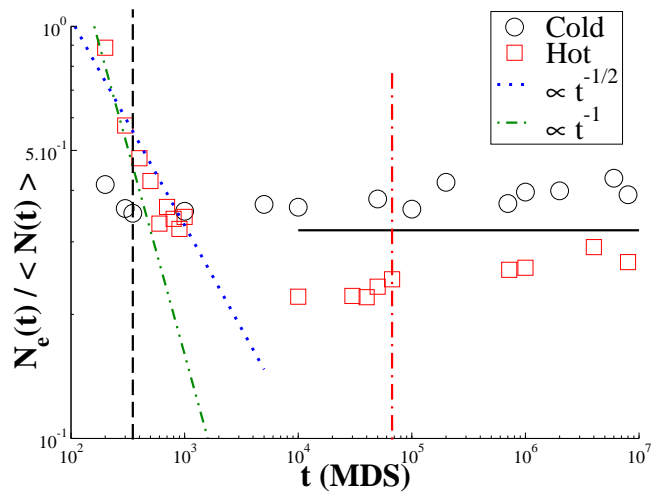

\rsfig{fig7_b.eps} \caption{(Color online) Ratio $N_e(t)/\langle
N(t)\rangle$ for polydisperse systems \emph{Cold} and \emph{Hot}
against simulation time. Dashed (mid-dashed) vertical line separates
generation to relaxation stages for the \emph{Cold} (\emph{Hot})
method. Also shown is the same ratio from Sukumaran\cite{Sukumaran}
for chains length $N=200$ as an indicative value.}
\label{ppa_poly.fig}
\end{figure}

The PPA analysis has been performed at different simulation times
(during generation and relaxation stages) and results are shown in
figures ~\ref{ppa_homo.fig} and \ref{ppa_poly.fig}.

Fig.~\ref{ppa_homo.fig} displays the number of monomers in straight
primitive path segments $N_{pp}=N_e$ for FPO and \emph{HotMono}
($n_{MDSbG}=10$ and $n_{MDSbg}=300$) generation methods. The
vertical dashed line separates the generation and growth regimes.
The horizontal line is the entanglement length $N_e$ from
Sukumaran\cite{Sukumaran}, that is in excellent agreement with our
data. This asymptotic value is even reached during the generation
stage for the \emph{HotMono} technique with $n_{MDSbg}=300$: the
relaxation stage is not required for this system!

The \emph{radical-like method} appears to perform perform
particularly  well at the relaxation stage: the entanglement length
do not deviate much from the asymptotic value for the \emph{HotMono}
system in comparison to the FPO method.

The PPA analysis has also been implemented for  polydisperse
\emph{Hot} and \emph{Cold} systems. Fig.~\ref{ppa_poly.fig} shows
the ratio $N_e(t)/\langle N(t)\rangle$ for polydisperse systems
(\emph{Cold} and \emph{Hot}) against simulation time. During the
generation stage, the time scale is given in $n_{growth}$ steps
units, whereas given in $n_{relax}$ MD steps during the relaxation
stage.

For \emph{Cold} system, generation/relaxation transition is
represented by a dashed vertical line, while a dot-dashed line is
used for \emph{Hot} system.

The same indicative value for the entanglement length $N_e / N$ from
Sukumaran\cite{Sukumaran} for chain length of size $N=200$ is also
shown, and must be considered as a mean value for both polydisperse
systems.

Indeed, the mean chain length at the end of the generation phase for
\emph{Cold} system is $\langle N \rangle_{\textit{Cold}}
(t\rightarrow\infty) = 172$, while for the \emph{Hot} $\langle
N\rangle_{\textit{Hot}}(t\rightarrow\infty)= 226$ (see table
~\ref{TAB_SIMU}).

For the \emph{Cold} method, the investigated ratio is almost
constant along the whole relaxation stage, during which
entanglements do not vary much.

For the \emph{Hot} system, this ratio displays a more complex
behavior. First, a power  law decrease, as noted by dotted ($\propto
t^{-1/2}$) and dotted-dashed line ($\propto t^{-1}$), is observed,
until $n_{growth}\sim 700$, corresponding to a ratio $N_e(t)/\langle
N(t)\rangle\sim 1/3$ nearly equal to results from \cite{Sukumaran}
for $N=200$ homopolymer chain melts.

In this regime, $\langle N(t)\rangle$ grows more rapidly than
$N_e(t)$, and the  growth process of each chain interacts with  a
stochastic background associated with the ensemble of  growing
chains. Thus, in this Rouse-like regime, topological constraints do
not play a significant role and one may expect that chains with
average length $\langle N(t)\rangle < N_e \sim N/3$ dominate the
polymerization, following a Rouse-like chain dynamics.

Following this regime, while $\langle N(t)\rangle$ still grows, a
stabilization of the same ratio is observed. In this regime, $N_e(t)
/ \langle N(t)\rangle < 1/3$, and a slowing down is observed during
chain growth dynamics. This new reptation-like regime, corresponds
to a dynamics where the surrounding medium topology limits
transverse chains displacements around their own contour length.
Chains with mean size $\langle N(t)\rangle > N_e \sim N/3$ follow
this reptation-like dynamics, and the polymerization process is
slowed down. While the longest chains are still growing, the average
entanglement length does not vary drastically, as one can see once
the generation stage is finished, where the ratio $N_e(t)/\langle
N(t)\rangle\rightarrow N_e/N$.

From all these results,  it appears that our approach is validated
as a method for generating equilibrated configurations  of
homopolymer melts. In the following section, the \emph{radical-like}
algorithm will be used to generate block copolymers in a lamellar
configuration.

\section{Application to copolymer generation}
\label{sec:copo}

In this section, generation of block copolymers will be investigated
and our \emph{radical-like} polymerization algorithm will be
modified to get a lamellar structure.

Modeling the demixion itself is not an easy task. Adjusting
force-fields and replicating basic units of previously assembled
copolymers, Srinivas\cite{Srinivas} managed to obtain large scale
demixtion in biological systems (self-assembled copolymers in
water). Zang\cite{Zhang} used full-atomistic simulations based on
dynamics density functional theory but their approach is limited to
small system sizes. May be more adapted to block copolymer
generation, semi-particle based methods such as Single Chain in
Mean-field~\cite{Muller_2,Daoulas,Sides,Daoulas_2} seem to be
promising.

In the following, we present an alternative method based on an
adaptation of the radical-like  method to the particular case of
symetric AB diblock where $N_A=N_B$ and $N=N_A+N_B$. $L_x$, $L_y$
and $L_z$ are the box sizes along the $x$, $y$ and $z$ directions.

\begin{figure}
\rrsfig{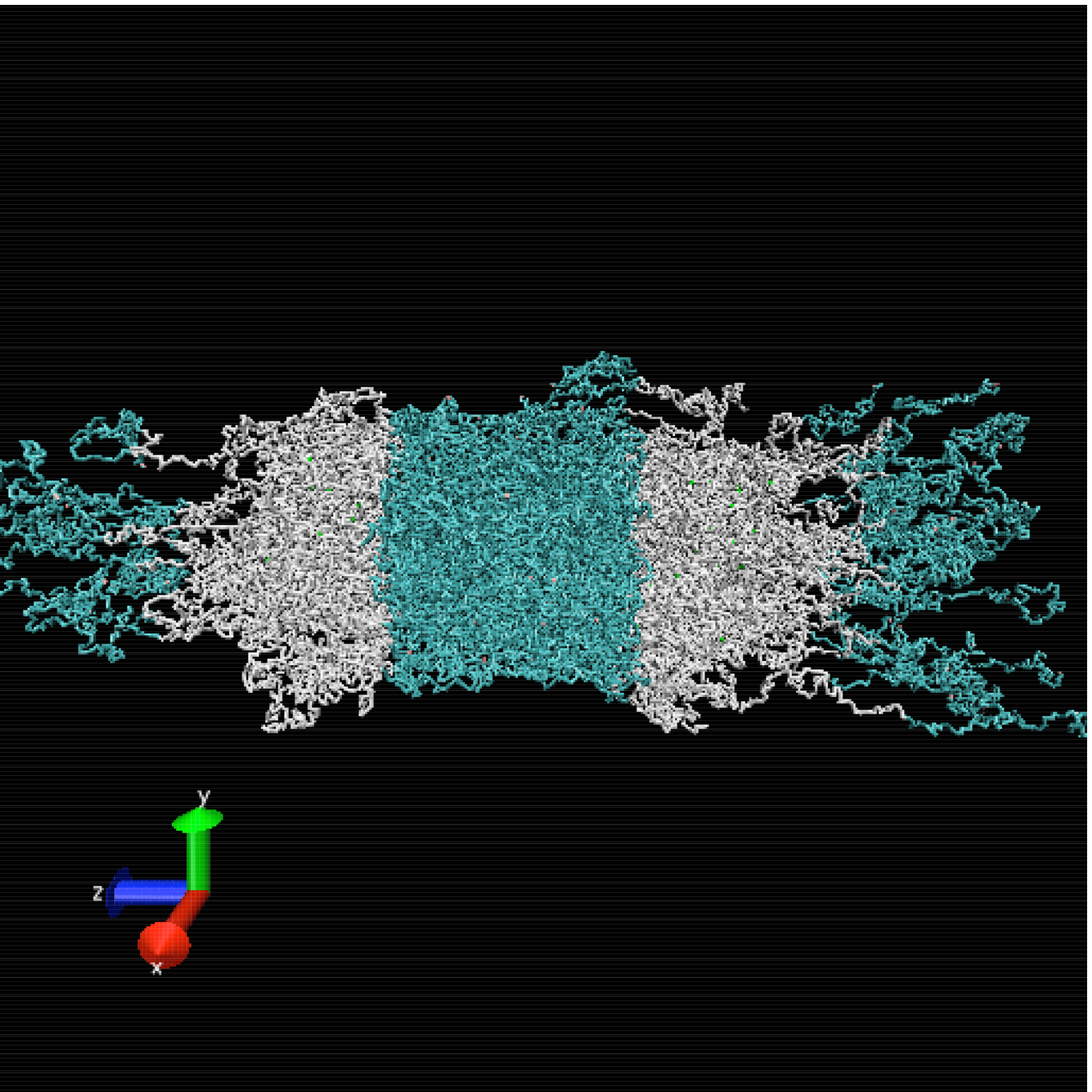}{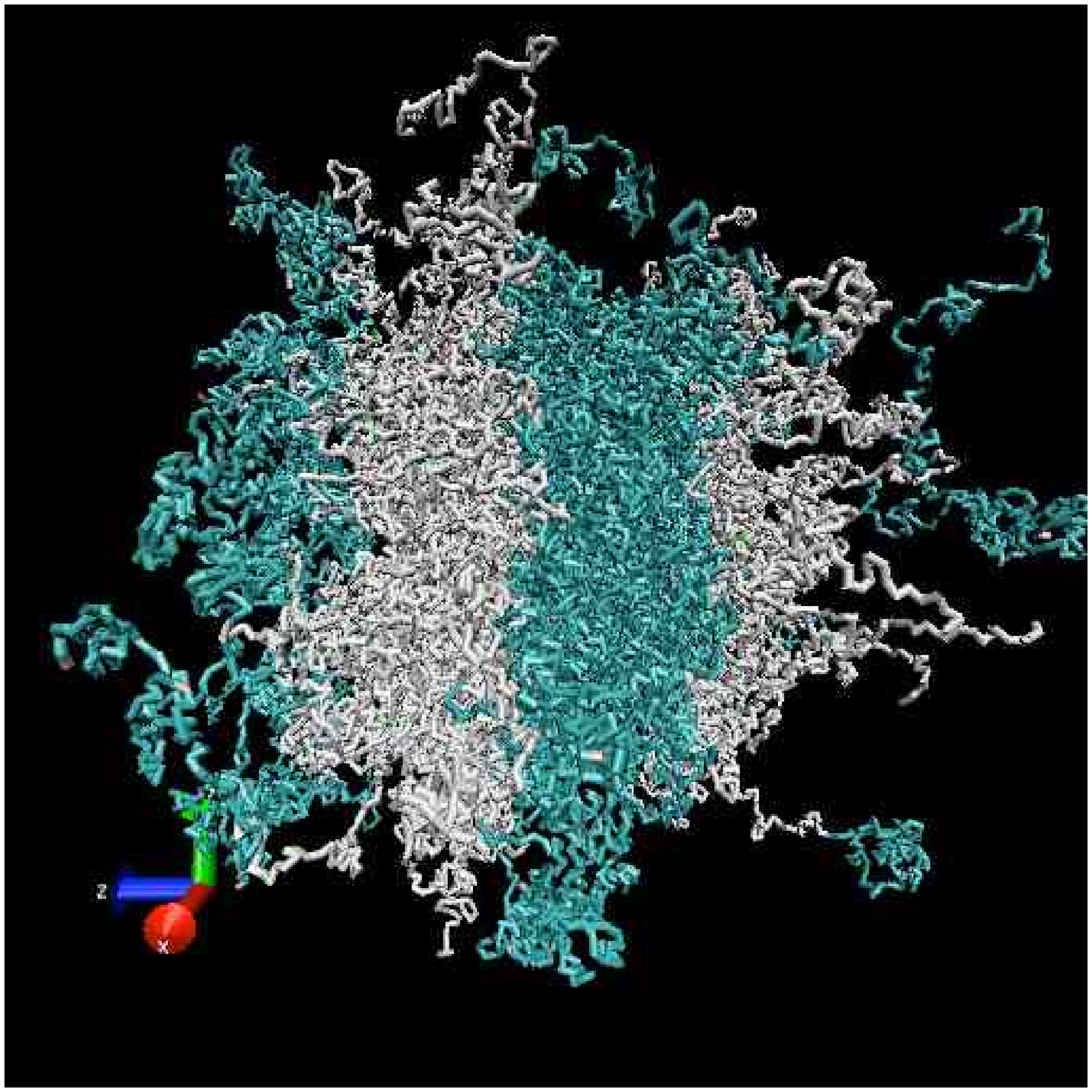} \caption{\label{dibloc_vs_chiN}
(Color online) Snapshots of symmetric di-block copolymers generated
using the \emph{radical-like} co-polymerization method for $M=215$
chains of length $N=200$, and for two values of the Flory-Huggins
compatibility parameter $\chi N=396$ (segregation regime -
\textbf{up}), and $\chi N=4$ (one phase region: mixing in progress -
\textbf{down}).}
\end{figure}

Generation of a di-block copolymer with an interface lying in the
($xy$) plane is performed as follows, starting from a Lennard-Jones liquid of monomers:

\begin{enumerate}
\item \textbf{Nucleation:} Each monomer $i$ has a probability $p$ to
be a radical of type $A$ if, say, $z_i>L_z/2$ and B otherwise.
\item \textbf{Growth:} As long as the chain does not reach the size
 $N/2$ ($N(t)<N/2$), growth is performed as in a homopolymer with a supplementary condition:
 addition of a new monomer $j$ is possible only if it lies in the
 same region ($z_j>L_z/2$ for $A$ chains and $z_j<L_z/2$ for $B$ chains).
 The interface situated at $z=L_z/2$ is then impermeable: no chain can cross it.
\item \textbf{From one region to the other:} Once a chain reach the critical size
$N/2$ ($N(t)=N/2$, the growth within a lamella is stopped. A force
is applied to attract the chain ends to the interface, and the
condition above is reversed: addition of a new monomer
 $j$ is possible only if it lies in the opposite region ($z_j<L_z/2$ for $A$ chains
 and $z_j>L_z/2$ for $B$ chains).Under this new condition, and once a radical
 combines with
 a new monomer in the opposite region, it turns into the opposite species ($A$ radical
 becomes $B$ radical and $B$ radical becomes $A$ radical). For chains,
 with
 $N(t)>N/2$, the growth is then continued with the impermeable interface condition:
 addition of a new monomer $j$ is possible only if it lies in the same region ($z_j>L_z/2$
 for $A$ chains and $z_j<L_z/2$ for $B$ chains). Growth a a chain occurs until its length reach the size $N$.
\item \textbf{Relaxation:} As for homopolymers, a number $n_{MDSbg}$ of MD steps is
performed between each growth step, during which the systems is coupled
to the heat bath at $k_B T=2\epsilon_{\alpha\alpha}$ and $P=0.5$.
\end{enumerate}

With this procedure, all chains have the same length $N=N_A + N_B$
and $N_A=N_B$. Systems are then relaxed at $k_B
T=0.5\epsilon_{\alpha\alpha}$ and $P=0.5$ during $10^6$ MD steps.

Values for excluded volume potentials ~\ref{LJpot} and
~\ref{FENEpot} have been chosen as,
$\epsilon_{AA}=\epsilon_{BB}=1.0$ and the interaction with the
solvent fixed to $\epsilon_{\alpha s}=1.0$ where $\alpha\in[A,B]$
and $s$ being the solvent molecular type. All
$\sigma_{\alpha\beta}=1.0$ while potentials are truncated and
shifted at $r_c=2.5\sigma_{\alpha\beta}$.

The order-disorder transition temperature is governed by the product
of $\chi N$, where $\chi= (\epsilon_{AA} + \epsilon_{BB} -
2\epsilon_{AB})/(2k_B T)$ is the Flory-Huggins temperature-dependant
interaction parameter characterizing the AB incompatibility.
Symmetric diblock copolymers are homogeneous at small $\chi N$
value, but strongly heterogeneous with ordered structure when $\chi
N$ exceeds, in mean-field theory, the critical order-disorder
transition value $\chi N_{ODT}\approx 10$. Hence, as a first
application of our \emph{radical-like copolymerization} algorithm,
we simulated such diblock copolymers in the two limiting case of
weak segregation limit with $\chi N=4$ (($\epsilon_{AB}=0.99$), and
the strong one, with $\chi N=396$ ($\epsilon_{AB}=0.01$).

To observe box dilatation and inter-lamellae distance relaxation, an
anisotropic  Nos\'e-Hoover barostat has been used during $5.10^6$ MD
steps, in such a way that $P_x=P_y=P_z=0$, while the temperature was
fixed to $k_B T = 0.5 \epsilon_{\alpha\alpha}$.

Snapshots of diblock configurations are shown in
Fig.~\ref{dibloc_vs_chiN}, where simulations have been performed on
$M=215$ chains with a polymerization degree of $N=200$. In  the
upper frame of Fig.~\ref{dibloc_vs_chiN}, one can observe that the
interface separating the two blocks is well defined and stable, as
expected in the strong segregation limit. On the contrary, in the right panel, the same interface is poorly defined and appears to be unstable on the simulation timescale.  One expects the diblock to become homogeneous in the long time limit.

This preliminary study on diblock copolymers allowed us to validate
the \emph{radical-like} copolymerization technics. The advantage of
this technique  resides in the control of the geometry of simulated
copolymers, as well as the possibility to generate in a flexible way
configurations with various topologies and chain architectures.

\section{Conclusion}
\label{sec:conclusion}

\begin{enumerate}
\item the \emph{radical-like} polymer chains generation
method is inspired by radical polymerisation in which
the reactive center of a polymer chain consists of a radical. The free radical reaction
mechanism can be divided in to three stages: (i) initiation (creation of free radicals);
(ii) propagation (construction of the repeating chain): (iii) termination (radical is no longer active).

\item Performing an relatively important number of
MD relaxation steps between each growth step (typically 300) leads to well equilibrated
chains (in terms of gyration radii, Mean Square Internal Distance, and Primitive Path Analysis),
 even for relatively short relaxation stage ($10^6$ MD Steps).

\item The main advantage of the radical-like generation algorithm is that equilibration
occurs simultaneously as chain growth, within a coarse grained molecular dynamics scheme.

\item Polymer melts generated with the radical-like algorithm are as well equilibrated
as melts generated by more classical methods (like fast push-off).

\item The radical-like generation method is particularly adapted to generate
polydisperse polymer melts (branched polymers, star polymers, copolymers,...).

\item nano-structured lamellar block copolymers have been successfully generated with the radical-like method.

\item physical and mechanical properties of di-block and tri-block copolymers generated using this algorithm,
will be the subject of a futur paper.
\end{enumerate}

\begin{acknowledgments}
  During the course of this work, we had valuable discussions with  R.~Estevez and D.~Brown.
  Computational support by IDRIS/France, CINES/France and the Federation Lyonnaise de Calcul Haute Performance
  is also acknowledged. Part of the simulations were carried out using the LAMMPS molecular dynamics software
  (http://lammps.sandia.gov). Financial support from ANR Nanomeca is also acknowledged.
\end{acknowledgments}

\end{document}